\documentclass[aps,prb,twocolumn,showpacs,preprintnumbers,amsmath,amssymb,superscriptaddress]{revtex4}%

\usepackage{graphicx}%
\usepackage{dcolumn}
\usepackage{amsmath}
\usepackage{color}
\usepackage{ulem}
\usepackage{multirow}

\begin{document}

\title{Pressure induced magnetic order in FeSe}

\author{Rustem~Khasanov}
 \email{rustem.khasanov@psi.ch}
 \affiliation{Laboratory for Muon Spin Spectroscopy, Paul Scherrer Institut, 5232 Villigen PSI, Switzerland}
\author{Zurab~Guguchia}
 \affiliation{Laboratory for Muon Spin Spectroscopy, Paul Scherrer Institut, 5232 Villigen PSI, Switzerland}
\author{Alex~Amato}
 \affiliation{Laboratory for Muon Spin Spectroscopy, Paul Scherrer Institut, 5232 Villigen PSI, Switzerland}
\author{Elvezio~Morenzoni}
 \affiliation{Laboratory for Muon Spin Spectroscopy, Paul Scherrer Institut, 5232 Villigen PSI, Switzerland}
\author{Xiaoli~Dong}
 \affiliation{Beijing National Laboratory for Condensed Matter Physics, Institute of Physics {\rm \&} University of Chinese Academy of Sciences, CAS, Beijing 100190, China}
\author{Fang Zhou}
 \affiliation{Beijing National Laboratory for Condensed Matter Physics, Institute of Physics {\rm \&} University of Chinese Academy of Sciences, CAS, Beijing 100190, China}
\author{Zhongxian~Zhao}
 \affiliation{Beijing National Laboratory for Condensed Matter Physics, Institute of Physics {\rm \&} University of Chinese Academy of Sciences, CAS, Beijing 100190, China}

\begin{abstract}
The magnetic order induced by the pressure was studied in single crystalline FeSe by means of muon-spin rotation ($\mu$SR) technique. By following the evolution of the oscillatory part of the  $\mu$SR signal as a function of angle between the initial muon-spin polarization and 101 axis of studied crystal it was found that the pressure induced magnetic order in FeSe corresponds either to the collinear (single-stripe) antiferromagnetic order as observed in parent compounds of various FeAs-based superconductors or to the Bi-Collinear order as obtained in FeTe system, but  with the Fe spins turned by 45$^{\rm o}$. The value of the magnetic moment per Fe atom was estimated to be $\simeq 0.13-0.14$~$\mu_{\rm B}$ at $p\simeq 1.9$~GPa.
\end{abstract}
\maketitle

Since their discovery in 2008,\cite{Hsu_PNAS_2008} iron chalcogenide  superconductors has attracted much interest. Being composed of a single layer of square Fe lattice tetrahedrally coordinated by chalcogene (Ch) atoms (Ch= Se, Te, S), FeCh consists just of a fundamental building blocks of Fe-based high-temperature superconductors (Fe-HTS).
The iron selenide, FeSe, superconducts at ambient conditions with the transition temperature $T_c\simeq8$~K.
Early muon spin rotation ($\mu$SR) experiments on FeSe revealed that the system is non-magnetic at ambient pressure
down to $T\simeq  0.02$~K.\cite{Khasanov_PRBR_2008} The first pressure experiments also do not detect the magnetic order up to pressures at least $p\sim20$~GPa.\cite{Medvedev_NatM_2009} This is in striking contrast to the other Fe-HTSs that usually exhibit static magnetic order in the parent compound.\cite{Buchner_NatM_2009, Dai_RMP_2015}
%
Shortly after, the NMR studies showed a wipeout of the signal that revealed an incipient magnetic phase transition under pressure.\cite{Imai_PRL_2009} It was further realized that pressure promotes the static magnetism in FeSe which was confirmed in $\mu$SR experiments by Bendele and coworkers.\cite{Bendele_PRL_2010, Bendele_PRB_2012} The static magnetic orders which competes with superconductivity was set in above $p\simeq 0.8$~Gpa  and both ground states were found to coexist on an atomic length for pressures exceeding $\sim 1.2$~GPa.

So far the only confirmation of pressure induced bulk magnetic order in FeSe was obtained from $\mu$SR data.\cite{Bendele_PRL_2010, Bendele_PRB_2012} Only very recently the appearance of bulk magnetism in high-quality FeSe single crystalline samples was confirmed in M\"{o}ssbauer under pressure experiments by Kothapalli {\it et al.}\cite{Kothapalli_Arxiv_2016} and NMR studies of Wang {\it et al.}\cite{Wang_Arxiv_2016}
%
It should be emphasized, however, that the exact magnetic-spin arrangement in FeSe is still unknown. The problem stems from the low value of the ordered magnetic moment on the Fe site ($m_{\rm Fe}$). Following $\mu$SR studies, for pressures $p\lesssim 2.5$~GPa $m_{\rm Fe}$ do not exceed $0.2$~$\mu_{\rm B}$.\cite{Bendele_PRB_2012} Recent M\"{o}ssbauer experiment results in similar estimate of $m_{\rm Fe}$ for $p\lesssim $4~GPa.\cite{Kothapalli_Arxiv_2016} Such small values of $m_{\rm Fe}$ makes the determination of the magnetic structure by means of neutron experiments to be quite challenging. We are only aware of one neutron diffraction measurement allowing to set the upper limit of $m_{\rm Fe}< 0.5-0.7$~$\mu_{\rm B}$ for pressures $p\lesssim$4.5~GPa.\cite{Bendele_PRB_2012}

\begin{figure}[htb]
\includegraphics[width=1.0\linewidth]{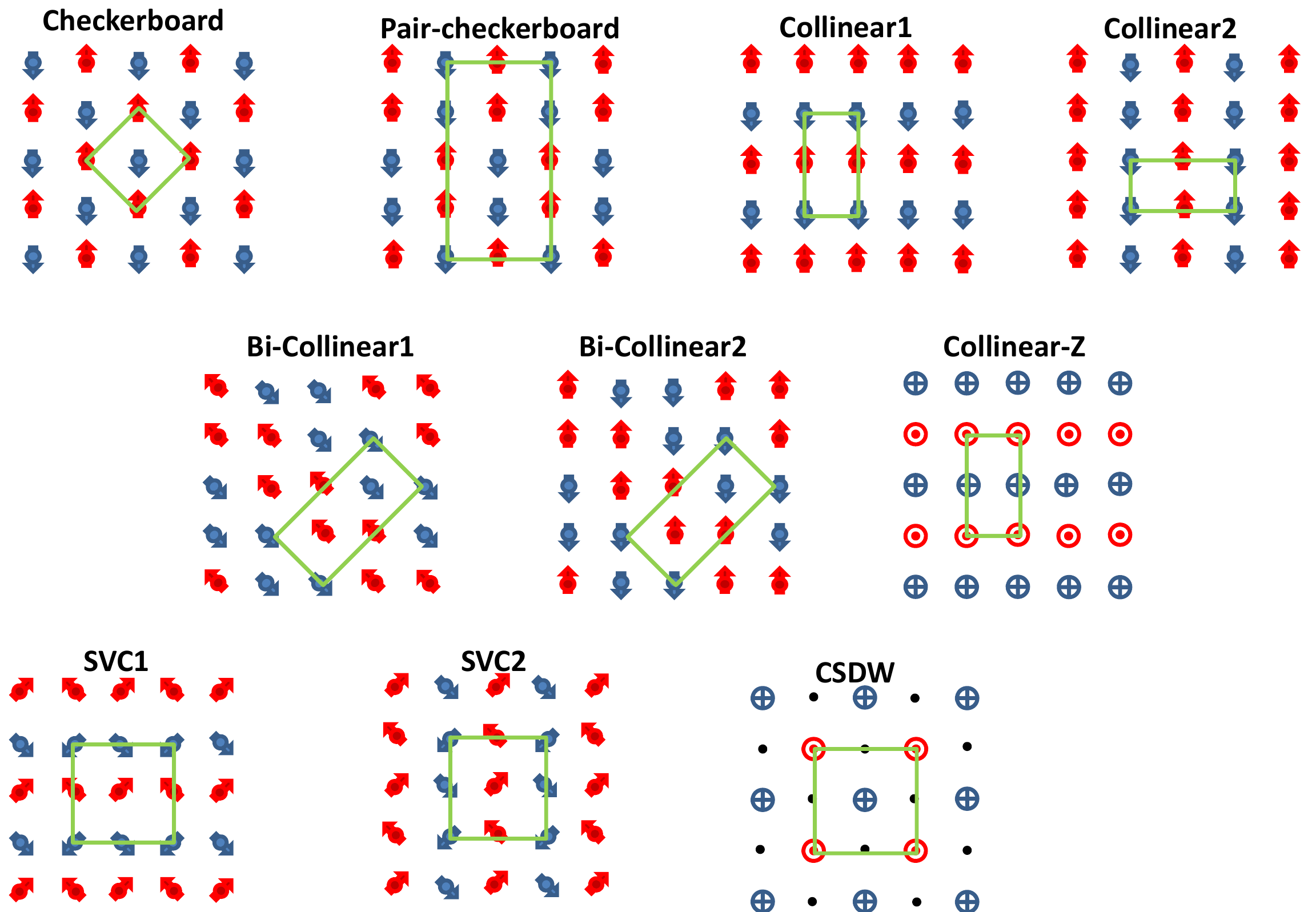}
%
\caption{The in-plane view of ten magnetic phases tested in the present study. The red and the blue arrows(circles) represent Fe spins. Green rectangles denote the in-plane magnetic unit cell. SVC1, SVC2, and CSDW phases preserves the tetragonal symmetry of the system, while the rest of the structures could be observed in both, the orthorhombic and the tetragonal magnetic unit cells.  }
 \label{fig:Structures}
\end{figure}

Experimentally, the antiferromagnetic (AFM) collinear structure consistent of stripes of parallel spins (the 'Collinear1` structure, Fig.~\ref{fig:Structures}) was established  for parent compounds of FeAs-based Fe-HTSs.\cite{Lynn_PhysC_2009, Dai_RMP_2015} The so-called 'Bi-collinear` order (denoted as 'Bi-Collinear1`, Fig.~\ref{fig:Structures}) was resolved for parent compounds of iron tellurides.\cite{Bao_PRL_2009, Li_PRB_2009, Dai_RMP_2015} For FeSe the direct measurements are still missing and only theoretical considerations were made till now.  For bulk FeSe the 'Collinear2` type of order was proposed in Ref.~\onlinecite{Ma_PRL_2008}  based on first principle calculations. The AFM 'Pair-checkerboard` order in bulk and mono-layer FeSe was considered by Cao {\it et al.}\cite{Cao_PRBR_2015}  For the mono-layer and the bi-layer FeSe films the 'Checkerboard` and the mixture of the ''Checkerboard` and 'Collinear1` orders were predicted in Ref.~\onlinecite{Liu_CPL_2015}.  The purely 'Pair-Checkerboard' order in bulk and thin-film FeSe was obtained in Refs.~\onlinecite{Tresca_2DM_2015}. The authors of Ref.~\onlinecite{Wang_arxiv_2015} have reported that the 'Pair-Checkerboard` and the 'Collinear1' stripe orders are realized in thin-film and bulk FeSe, respectively.

The aim of this paper was to identify the type of the pressure induced magnetic order in FeSe single crystalline sample by means of muon-spin rotation technique. In addition to the mentioned above, the 'Bi-Collinear2` order which differs from the 'Bi-Collinear1` one by spins turned to 45$^{\rm o}$ (see Fig.~\ref{fig:Structures} and Refs.~\onlinecite{Tresca_2DM_2015, Liu_CPL_2015}) as well as several magnetic structures suggested by Christensen {\it et al.}\cite{Christensen_PRB_2015} for various electron- and hole-doped Fe-HTSs were considered. This includes the collinear order denoted as 'Collinear-Z`, two type of spin vortex crystal (SVC) phases and the charge-spin density wave (CSDW) phase. Note that SVC1, SVC2, and CSDW phases preserves the tetragonal symmetry of the system, while the rest of the structures could be observed in both, the orthorhombic and the tetragonal magnetic unit cells.  By following the evolution of the oscillatory part of the  $\mu$SR signal as a function of angle between the initial muon-spin polarization and 101 axis of studied crystal it was shown that the pressure induced magnetic order in FeSe may correspond either to the  ''Collinear1``  or the 'Bi-Collinear2` type of the order (see Fig.~\ref{fig:Structures}).

The FeSe single crystal was synthesized by means of floating zone technique as is described in Ref.~\onlinecite{Ma_SST_2014} The x-ray experiment confirms that crystal exhibit a single preferred orientation of a tetragonal (101) plane.\cite{Ma_SST_2014}
The muon-spin rotation ($\mu$SR) experiments were carried out at the $\mu$E1 beam line by using the dedicated GPD spectrometer (Paul Scherrer Institute, Switzerland). The zero-field (ZF) and 3~mT weak transverse-field (TF) $\mu$SR measurements were performed at temperatures ranging from $\simeq$5 to 60~K and pressure $p\simeq 1.9$~GPa. The typical counting statistics were $\sim20\cdot 10^{6}$ positron events for each data point. The experimental data were analyzed by using the MUSRFIT package.\cite{Suter_MUSRFIT_2012}
The pressure was generated in a double-wall piston-cylinder type of cell made of MP35N alloy. As a pressure transmitting medium 7373 Daphne oil was used. The details of the experimental setup for conducting $\mu$SR under pressure experiments are given in Ref.~\onlinecite{Khasanov_HPR_2016}.

\begin{figure}[t]
\includegraphics[width=1.0\linewidth]{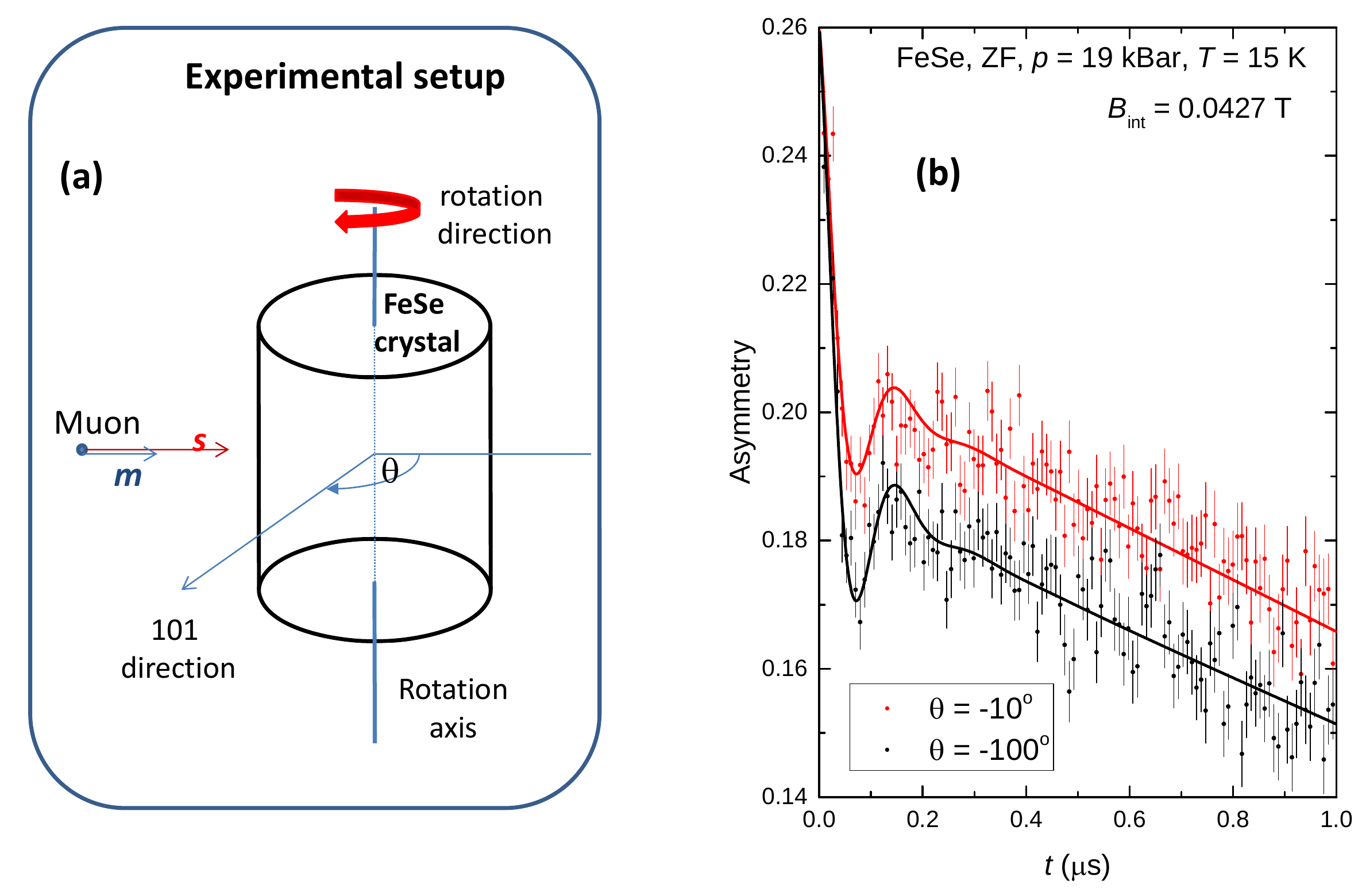}
%
\caption{(a) The schematic of the experiment. The rotation of cylindrically shaped FeSe crystal placed inside the pressure cell allows to change the angle $\theta$ between the initial muon-spin polarization and 101 crystallographic axis. (b) ZF-$\mu$SR time spectra measured at $T=15$~K, $p=1.9$~GPa for $\theta=-10^{\rm o}$ (red symbols) and $-100^{\rm o}$ (black symbols). The solid lines are fits by using Eq.~\ref{eq:ZF_Asymmetry_PC}. }
 \label{fig:experiment_and_ZF}
\end{figure}

Figure \ref{fig:experiment_and_ZF}~a shows the schematic of the experiment. The sample, placed inside the pressure cell, was rotated in a way allowing to change the angle $\theta$ between the initial muon-spin polarization {\it P}(0) and 101 crystallographic axis of the sample. Typical zero-field muon time spectra measured at $T=15$~K and $p\simeq 1.9$~GPa for $\theta=-10^{\rm o}$ and $-100^{\rm o}$ are shown in Fig.~\ref{fig:experiment_and_ZF}. The solid lines are fits of the following function to the experimental data:
\begin{equation}
A(t)=A_s(0) P_s(t)+ A_{pc}(0) P_{pc}(t),
 \label{eq:ZF_Asymmetry_PC}
\end{equation}
Here $A_{s}(0)$ and $A_{pc}(0)$ are the initial asymmetries and $P_{s}(t)$ and $P_{pc}(t)$ are the muon-spin polarizations belonging to the sample and the pressure cell, respectively. The polarization of the pressure cell $P_{pc}(t)$ was obtained in separated set of experiments.\cite{Khasanov_HPR_2016}
The polarization of the sample was described by the following functional form:
\begin{equation}
P_{s}(t)=f_{\rm osc}e^{-\lambda_T t} \cos(\gamma_\mu
B_{\rm int}t) + (1-f_{\rm osc})\;e^{-\lambda_L t}.
 \label{eq:P_FM}
\end{equation}
Here $B_{\rm int}$ is the internal field on the muon stopping site, $\gamma_{\mu}= 2\pi\;135.5$~MHz/T is the muon gyromagnetic ratio, and $\lambda_T$ and $\lambda_L$ are the transverse and the longitudinal exponential relaxation rates, respectively. The oscillating ($f_{\rm osc}$) and  non oscillating ($1-f_{\rm osc}$) $\mu$SR signal fractions originate from the magnetic field components which are transversal to the initial muon-spin polarization and cause a precession [$B_{\rm int}\perp P(0)$], and the non-precessing longitudinal field components [$B_{\rm int}\parallel P(0)$], respectively. Note that in powder sample, where all possible angles between $B_{\rm int}$ and $P(0)$ are equally possible, $f_{\rm osc}\equiv 2/3$. In the  single crystalline sample the value of $f_{\rm osc}$ may vary from 1, in  $B_{\rm int}\perp P(0)$ case, to 0 for  $B_{\rm int}\parallel P(0)$.

From the experimental data presented in Fig.~\ref{fig:experiment_and_ZF}~b two important points emerge:
(i) The spontaneous muon-spin precession with $B_{\rm int}\simeq 42.7$~mT is clearly detected  on the ZF $\mu$SR time spectra. Consequently, the static magnetic order in FeSe crystal studied here is established below the N\'{e}el temperature $T_N$ in agreement with the results of previous $\mu$SR experiments on FeSe powders,\cite{Bendele_PRL_2010, Bendele_PRB_2012} and recent   M\"{o}ssbauer experiments of high-quality single crystals.\cite{Kothapalli_Arxiv_2016}
(ii) The value of the oscillatory component $f_{\rm osc}$ depends on the angle between $P(0)$ and 101 axis of the crystal. $f_{\rm osc}$ for $\theta =-10^{\rm o}$ is obviously smaller than that for $\theta=-100^{\rm o}$. Consequently, by measuring $f_{\rm osc}$ as a function of $\theta$, the direction of the internal field $B_{\rm int}$ on the muon stopping position might be determined.

\begin{figure}[t]
\includegraphics[width=1.0\linewidth]{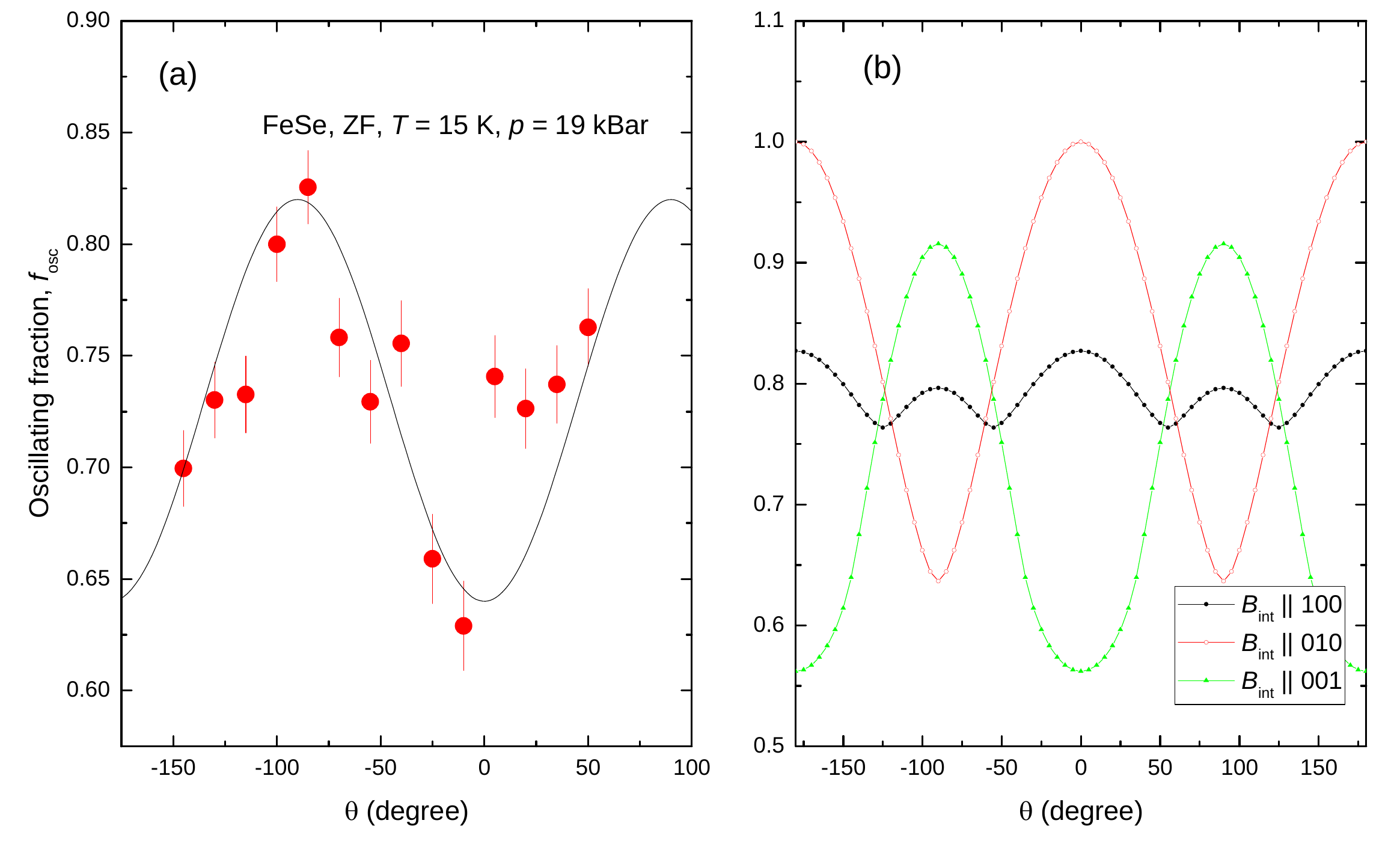}
%
\caption{ (a) Angular dependence of the oscillating fraction $f_{\rm osc}$ in FeSe single crystalline sample at $T=15$~K and $p=1.9$~GPa. (b) $f_{\rm osc}(\theta)$ calculated for the internal field $B_{\rm int}$ aligned along 100 (black), 010 (red) and 001 (green) crystallographic directions (see the Supplemental Information part for details). }
 \label{fig:fosc-theta}
\end{figure}

The dependence of $f_{\rm osc}$ on $\theta$ is shown in Fig.~\ref{fig:fosc-theta}~a. It has $180^{\rm o}$ periodicity with the maximum ($f_{\rm osc}^{max}$) and minimum ($f_{\rm osc}^{min}$) corresponding to $\theta=90^{\rm o}+n\cdot 180^{\rm o}$ and $n\cdot 180^{\rm o}$ ($n$ is the integer number), respectively. Bendele {\it et al.}\cite{Bendele_PRB_2012} have considered the collinear (single-stripe) AFM order in FeSe ('Collinear1`, Fig.~\ref{fig:Structures}) and shown that in such case $B_{\rm int}$ is aligned along the crystallographic $c-$axis.  The comparison of $f_{\rm osc}(\theta)$ with that obtained theoretically for $B_{\rm int}\parallel  100$, 010, and 001 crystallographic directions (Fig.~\ref{fig:fosc-theta}~b and the Supplemental part) is consistent with this statement. Indeed, for $B_{\rm int}\parallel 001$ the period, the values of $\theta$ corresponding to the minimum and maximum  of $f_{\rm osc}(\theta)$ are just the same as they observed experimentally. The theoretically calculated $f_{\rm osc}^{max}=0.92$ and $f_{\rm osc}^{min}=0.57$ are less than 10\% different from the experimentally obtained 0.83 and 0.62 (see Fig.~\ref{fig:fosc-theta}). Such small difference could be explained by some misalignment of 101 axis between different crystallites. As illustrated in Ref.~\onlinecite{Ma_SST_2014}, FeSe crystals prepared similarly to the one used in our studies are characterized by a high level of mosaicity  which results in a strong broadening of Bragg reflection peaks.
%

\begin{figure}[t]
\includegraphics[width=1.0\linewidth]{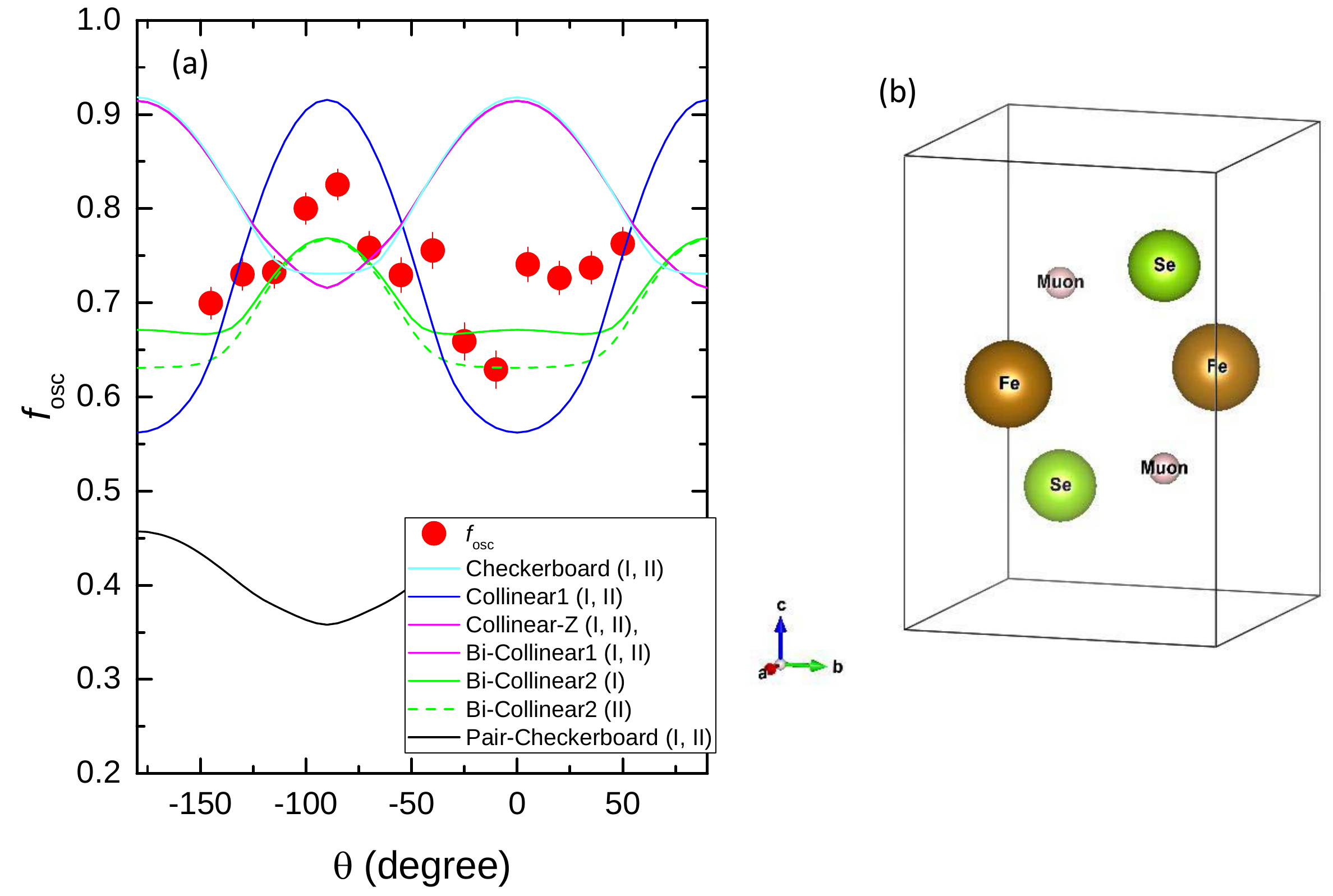}
%
\caption{ (a) $f_{\rm osc}$ as a function of $\theta$ calculated for 'Checkerboard`, 'Collinear1`, 'Bi-Collinear1`, 'Bi-Collinear2` and 'Collinear-Z` structures. ''I`` and ''II`` correspond to the magnetic unit cell to be the same or doubled in comparison to the chemical one along the $c-$axis. (b) The FeSe unit cell (crystallographic group $P4/nmm$ 129, origin choice 2).\cite{Khasanov_NJP_2010} Muons are on the 2c [$(1/4,1/4,z)$, $z_\mu=0.84$] Wyckoff position. The structure is visualized by using VESTA package.\cite{Vesta} }
 \label{fig:Bint_Structures}
\end{figure}

The above obtained consistency between $f_{\rm osc}(\theta)$ and 'Collinear1` type of order do not allow us, however, to make any firm conclusion about other magnetic phases as they presented in Fig.~\ref{fig:Structures}.  To obtain more quantitative information, calculations of corresponding internal fields at muon stopping sites were carried out.
The muon sites in FeSe were previously calculated by Bendele {\it et. al.}\cite{Bendele_PRB_2012} by using the modified Thomas Fermi approach.  There are two equivalent minima in the unit cell  corresponding to the 2c [$(1/4,1/4,z)$, $z_\mu=0.84$; according to the crystallographic group $P4/nmm$ 129, origin choice 2] Wyckoff position (see Fig.~\ref{fig:Bint_Structures}~b).
The spontaneous local field for the muon site $i$ was assumed to be entirely determined by the dipolar component:
\begin{equation}
{\bf B}_{{\rm int},i}\simeq {\bf B}_{{\rm dip},i},
 \label{eq:Bloc}
\end{equation}
with $B_{\rm dip}$ at position ${\bf r}$ within the lattice unit cell calculated via:\cite{Blundell_PhysB_2009, Maisuradze_PRB_2011}
\begin{equation}
B_{\rm dip}^\alpha({\bf r})=\frac{\mu_0}{4\pi} \sum_{j,\beta} \frac{m_j^\beta}{R_i^3}
 \left( \frac{3 R_j^\alpha R_i^\beta}{R_j^2} - \delta^{\alpha \beta} \right).
 \label{eq:B_dip}
\end{equation}
Here ${\bf R}_j={\bf r}-{\bf r}_j$, $\alpha$ and $\beta$ denote the vector components $x$, $y$, and $z$, ${\bf r}_j$ is the position of $j-$th magnetic ion in the unit cell, and $m_j^\beta$ is the corresponding magnetic moment. The summation is taken over a sufficiently large Lorentz sphere of radius $R_L$.
Note that Eq.~\ref{eq:Bloc} differs from its general form which includes also the so-called contact field term ${\bf B}_{\rm cont}=A_{\rm cont}\sum_{k=1}^N {\bf m}_k$ ($A_{\rm cont}$ is the contact constant and the summation is made over the  $N$ nearest neighboring magnetic moments).\cite{Amato_PRB_2014, Khasanov_PRB_2016} Thanks to the 2c  Wyckoff position of the muon in FeSe lattice (Fig.~\ref{fig:Bint_Structures}~b), the sum of  ${\bf m}_k$ becomes zero for all structures
presented in Fig.~\ref{fig:Structures} with the magnetic unit cell doubled in comparison with the chemical one along the crystallographic $c-$directions, and for all structures except both 'Bi-Collinear` ones for a case without doubling.

The results of internal field calculations for various magnetic structures presented in Fig.~\ref{fig:Structures} and $m_{\rm Fe}=1$~$\mu_{\rm B}$ are summarized in Table~1 in the Supplemental part. The part of the data denoted as ''I`` corresponds to the case when the magnetic and the chemical unit cell have similar $c-$axis constants (the magnetic order along the $c-$axis is ferromagnetic). For the part denoted as ''II`` the magnetic unit cell along the $c-$direction is doubled in comparison to the chemical one and  the magnetic order along the $c-$axis becomes antiferromagnetic. The last column shows the value of $m_{\rm Fe}$ as calculated from the experimentally obtained $B_{\rm int}\simeq 0.0427$~T (see Fig.~\ref{fig:experiment_and_ZF}~b).

Several magnetic phases could be excluded from the consideration based entirely on the dipolar field calculations. The 'Collinear2` structure results in zero internal field on the muon stopping position. The 'SVC1` and 'SVC2` result in two different $B_{\rm int}$'s. Both these findings are inconsistent with the experimentally observed single finite internal field value (see Fig.~\ref{fig:experiment_and_ZF}~b and Refs.~\onlinecite{Bendele_PRL_2010, Bendele_PRB_2012}). Values of $m_{\rm Fe}$ for 'CSDW` structure were estimated to be 0.70 and 1.04~$\mu_{\rm B}$ for the magnetic unit cell with and without doubling along the crystallographic $c-$direction, respectively. These values are bigger then the upper estimate of $m_{\rm Fe}< 0.5-0.7$~$\mu_{\rm B}$ set in neutron diffraction experiment.\cite{Bendele_PRB_2012}

Following the above discussion, our data are consistent with the internal field on the muon stopping position aligned along the crystallographic $c-$direction. Among the magnetic phases left, the 'Collinear1` phase, which was already proposed in Ref.~\onlinecite{Bendele_PRB_2012}, satisfy such criteria.  In order to check if the rest of the phases ('Checkerboard`, 'Pair-Checkerboard`, 'Bi-Collinear1`, 'Bicollinear2` and 'Collinear-Z`) could be consistent with the experiment, the corresponding $f_{\rm osc}(\theta)$ dependencies were calculated (Fig.~\ref{fig:Bint_Structures}~a). Since the AFM order in Fe-HTSs is generally
preceded by a tetragonal-to-orthorhombic lattice distortion,\cite{Dai_RMP_2015} the twinning effects (rotation of the structure within the $ab-$plane by 90$^{\rm o}$) were also considered. The results presented in Fig.~\ref{fig:Bint_Structures} imply that two type of orders, namely the 'Collinear1` and 'Bi-Collinear2` become consistent with the experiment.

Based entirely on the experimental data one can not distinguish between two suggested above magnetic structures. There are few arguments, however, in favor of 'Collinear1` rather than 'Bi-Collinear2` type of magnetic order.
(i) Strong commensurate spin fluctuations with an in-plane wave vector ${\bf q}=(\pi, 0)$ were observed recently in FeSe at ambient pressure.\cite{Wang_NatM_2015,Ma_Arxiv_2016} Note that the in-plane ${\bf q}=(\pi, 0)$ corresponds to the stripe-like 'Collinear1` type of magnetic order.
(ii) Recent {\it ab initio} calculations have attributed the absence of static magnetic
order at ambient pressure in FeSe to competition between  different magnetic ordering vectors and shown that the application
of pressure lifts this near degeneracy, leading to a $(\pi, 0)$ stripe order.\cite{Glasbrenner_NatPh_2015}
(iii) The $(\pi, 0)$ nature of the pressure-induced magnetic state in FeSe is supported by the Fermi surface reconstruction
reported in quantum oscillations experiments.\cite{Terashima_PRB_2016}
(iv) The x-ray diffraction experiments of Kothapalli {\it et al.}\cite{Kothapalli_Arxiv_2016} show that the magnetic order in FeSe breaks the tetragonal symmetry of the lattice in the same manner as the stripe-type magnetic order in the other iron-based materials.\cite{Lynn_PhysC_2009, Dai_RMP_2015}

To conclude, the magnetic order induced by the pressure was studied in single crystalline FeSe by means of muon-spin rotation. By following the evolution of the oscillatory part of the  $\mu$SR signal as a function of angle between the initial muon-spin polarization and 101 axis of studied crystal it was found that the pressure induced magnetic order in FeSe corresponds either to the collinear (single-stripe) antiferromagnetic order as observed in parent compounds of various FeAs-based superconductors or to the Bi-Collinear order as obtained in FeTe system, but  with the Fe spins turned by 45$^{\rm o}$. The value of the magnetic moment per Fe atom was estimated to be $\simeq 0.13-0.14$~$\mu_{\rm B}$ at $p\simeq 1.9$~GPa.

The work was performed at the Swiss Muon Source (S$\mu$S), Paul Scherrer Institute (PSI, Switzerland). RK acknowledges helpful discussions with Rafael Fernandes. The work of ZG was supported  by the Swiss National Science Foundation (SNF-Grant 200021-149486). The work of XD, FZ and ZZ was supported by the "National Key Research and Development Program of China (Grant 2016YFA0300301)", the "Strategic Priority Research Program (B)" of the Chinese Academy of Sciences (Grant XDB07020100) and the Natural Science Foundation of China (Grant 11574370).

\end{document}